\documentstyle[epsfig,alex,12pt]{article}
\newcommand{\dT}{\hbox{${\rm T}\kern -7pt \raise 2.5pt\hbox{\tiny$|$}\kern 5.5pt$}}
\newcommand{\barray}{\begin{array}{l}}
\newcommand{\earray}{\end{array}}

\newcommand{\uraro}[2]{\hbox{$\kern 3pt\raise -2pt \hbox{$\raro$}
 \kern -14pt \raise
+3.5pt\hbox{\tiny{$#1\raro #2$}}$}}

\newcommand{\itms}[1]{\item[[#1\kern -5pt]]}

\newcommand{\II}{{\bf I\kern -1pt I}}

\newcommand{\sC}
   {\hbox{\tiny \hbox{${\rm C} \kern -5.0pt \raise 1pt \hbox{\tiny$|$}\kern 7.5pt$}}}

\def\newtheorems{\newtheorem{theorem}{Theorem}[section]

                 \newtheorem{lemma}[theorem]{Lemma}
                 \newtheorem{defn}[theorem]{Definition}
                 \newtheorem{claim}[theorem]{Claim}

                 }
                 \newtheorem{problem}{Problem}[section]
                 \newtheorem{notation}{Notation}[section]

\newtheorems

\newcommand{\bd}{\begin{description}}
\newcommand{\ed}{\end{description}}
\newcommand{\ben}{\begin{enumerate}}
\newcommand{\een}{\end{enumerate}}

\newcommand{\ga}{\alpha}

\newcommand{\raro}{\rightarrow}

\newcommand{\comp}{\hbox{$<\kern -3pt >$}}
\newcommand{\ncomp}
                {\;\hbox{\hbox{/}\kern -9.5pt \hbox{$<\kern -3pt >$}}}

\newcommand{\meet}
               {\hbox{$\wedge \kern -5.75pt \raise 1.5pt \hbox{$.$}\,$}}
\newcommand{\Meet}
             {\hbox{$\bigwedge \kern -8pt \raise 0.75pt \hbox{$.$}\:$}}
\newcommand{\ld}
               {\hbox{$< \kern -6pt \raise 2pt \hbox{$.$}\,$}}
\newcommand{\sss}{\: \hbox{$
\underline{\hbox{$\subset$}}\kern -4pt\raise -2pt \hbox{$\tiny |$}  
$}\: }
\newcommand{\rraro}[2]{\hbox{$\kern 3pt\raise 2pt \hbox{$\raro$}
 \kern -14pt \raise
-3.5pt\hbox{\tiny{$#1\raro #2$}}$}}

\newcommand{\frc}{\hbox{$\parallel \kern -5.7pt \hbox{$-$}$}}

\newcommand{\nfrc}{\not \kern -5pt \frc}
\newcommand{\rest}{\vbox{\hbox{$\:\kern -2pt\mathbin{\vert\kern-3.1pt\lower-1pt
   \hbox{$\mathsurround=0pt\mathchar"0012$}\kern-4pt}\:$}}}

\newcommand{\pcntda}{\lower5pt\hbox{$\stackrel{\subset}{\neq}$}}
\newcommand{\pcntdb}{\lower5pt\hbox{$\stackrel{\supset}{\neq}$}}
\newcommand{\pcntdc}{\lower5pt\hbox{$\stackrel{\subseteq}{\neq}$}}

\newcommand{\surj}{\vbox{\hbox{$\longrightarrow $
                  \kern -22pt \hbox{\lower 2.5pt  \hbox{\tiny onto}}
                  \kern -16pt \hbox{\raise 5pt  \hbox{\tiny 1-1}}
                  \kern 3pt}}}
\newcommand{\uarrow}[2]{\vbox{\hbox{$\longrightarrow $
                  \kern -16pt \hbox{\raise 5pt  \hbox{\tiny $#1$}}
                  \kern 10pt }}}

\newcommand{\R}{\hbox{I\kern-.1500em \hbox{\sf R}}}
\newcommand{\sR}{\hbox{\tiny \hbox{I\kern-.1500em \hbox{\sf R}}}}

\newcommand{\Q}
   {\hbox{${\rm Q} \kern -7.5pt \raise 2pt \hbox{\tiny$|$}\kern 7.5pt$}}
\newcommand{\C}
   {\hbox{${\rm C} \kern -7.5pt \raise 2pt \hbox{\tiny$|$}\kern 7.5pt$}}
\newcommand{\Z}{\hbox{\sf Z\kern-0.720em\hbox{ Z}}}
\newcommand{\sZ}{\hbox{\tiny\hbox{ \sf Z\kern-0.720em\hbox{ Z}}}}

\newcommand{\qed}{\kern 5pt\vrule height8pt width6.5pt depth2pt}
\newcommand{\text}[1]{\mbox{#1}}

\newcommand{\N}{\hbox{I\kern-.1500em \hbox{\sf N}}}

\begin{document}
\title{Generalizations of the Hanoi Towers Problem}

\author{Sergey Benditkis ~~~~~ and ~~~~~ Il'ya Safro\\
  supervised by Prof. Daniel Berend}

\date{ }
\maketitle

\centerline{\large Department of Mathematics and Computer Science}
\centerline{\large Ben-Gurion University of the Negev}
\centerline{\large Beer-Sheva, Israel}
\centerline{\large {\tt \{safro,serge\}@cs.bgu.ac.il}}  

\vspace{2in}

\begin{abstract}
  
  Our theme bases on the classical Hanoi Towers Problem. In this paper we will define a new problem, permitting some positions, that were not legal in the classical problem. Our goal is to find an optimal (shortest possible) sequence of discs' moves. Besides that, we will research all versions of 3-pegs classical problem with some special constraints, when some types of moves are disallowed.
\end{abstract}

%\pagebreak

\section{Introduction}

\par The first version of this puzzle was marketed by Edouard Lucas in 1883 under the name ``The Towers of Hanoi''.
It consisted of three pegs (or towers) on a wooden base and eight rings or disks of different sizes, arranged on one of the pegs, the largest at the bottom and the others in decreasing size, so that the smallest was at the top.
It was introduced as a variant of the mythical ``Towers of Brahma'', which had real towers and sixty four gold rings. According to Budhist legend, Judgment day will arrive when the monks from the monastery in the Low Himalyas finish to move all disks to the target peg.
\\\\
Let us state precisely the problems that we attempt to solve in this paper.

\begin{problem} 
  \rm Given are $n$ disks of sizes $1,2,...,n$, arranged on a peg in this order with the largest at the bottom and the smallest on top and two initially empty pegs. Transfer all discs to one of the other pegs, using the minimal possible number of moves, under the following constraints :

\begin{enumerate}
\item only one disc may be moved at a time,
\item only a topmost disk may be moved,
\item a disc cannot be placed on a smaller one.
\end{enumerate}

\end{problem}

For the next problems we introduce some definitions :

\begin{notation}
  Let $H(V,E)$ be a directed graph of possible legal moves, where
  \begin{itemize}
  \item $V$ is a set of pegs (in our case $|V|=3$).
  \item $E$ is a set of edges of the type $<a,b>$, where $<a,b>\in E$ means that moving a disc from peg a to b is legal
  \end{itemize}
\end{notation}

\begin{defn}
  \rm A state is {\it standard} if 2 pegs are empty and the $n$ discs are arranged with the largest at the bottom and the smallest on top on the third peg, when no large disc can be placed on the smaller one.
\end{defn}

\begin{defn}
  \rm {\it Model } M is a pair $<R,n>$, where $R$ are constraints on disc's transferring and $n$ is a number of discs. The goal will be to move all discs from standard state on peg X to standard state on peg Y.
\end{defn}

\begin{defn}
  {\rm Given are 2 discs A and B on peg X. Without lost of generality, disc A is placed higher than disc B. We say that distance between discs A and B is a $C=size(A)-size(B)$.}
\end{defn}

\begin{defn}
  {\rm Distance of the model $M$ is a maximal permitted distance between two arbitrary discs under constraints $R$ of $M$.}
\end{defn}

\begin{problem}
  \rm Given are a graph of type $H(V,E)$ and a number of discs $n$. The goal is to find an optimal sequence of moves from one standard state to another.
\end{problem}

The following problem was formulated by Wood [1] and is still open.

\begin{problem} \label{pmain}
  \rm Given are a model $M$ with following definition of $R$ :
  
  \begin{itemize}
  \item only one disk at a time may be moved
  \item only a topmost disk may be moved
  \item $C>0$ is a distance of the model
  \item every move X $\rightarrow$ Y is permitted if 3 previous rules are fulfilled
  \end{itemize}

\end{problem}

In the next sections we will deal with all these problems.
%\pagebreak
\section{The classical optimal solution (Problem 1.1)}
We now describe the classical solution to the puzzle, which is optimal. To transfer N discs from peg 1 to peg 2, say, first move the top $N-1$ discs (recursively) from peg 1 to peg 3, using peg 2 as intermediate storage. Once this is completed move disc N from peg 1 to peg 2, and then finally move the $N-1$ discs from peg 3 to peg 2 (recursively), using peg 1 as intermediate storage. Letting $T_{N}$ be the number of moves in this solution, we see that $T_{1}=1$, and $T_{N}=2T_{N-1}+1$. By induction, we can easly prove, that $T_{N}=2^{N}-1$.

That the classical solution is indeed optimal and is not hard to see. To transfer N discs from one peg to another, we must at some point move disc N at least once. In order to move disc N, it must be alone on its peg, and some other peg must be empty; hence the remaining peg must contain the $N-1$ smaller discs. Finally after disc $N$ has been moved to peg 2 for the last time the remaining $N-1$ discs have to be transferred from one peg to another.
\\
\par{Hence, if $T'(N)$ denotes the minimal total number of moves required to tranfer N discs, then $T'(N) \geq 1+2T'(N-1)$. Since $T'(1)=1$, we see that $T'(N) \geq 2^{N}-1$, which proves our claim.}

%\pagebreak
\section{Directed graphs with three nodes}

Assume we have a strongly connected directed graph $G=(V,E)$ such that $|V|=3$.
The nodes of the graph are pegs (Let us mark them 1,2,3).
Assume that the $n$ discs are placed on peg 1 in the standard position.
\\
Our target is to transfer $n$ discs to peg 2, under the following constraints:
\begin{itemize}
  \item{All constraints from the classical problem,}
  \item{We can move a disc from peg $i$ to peg $j$ only if $(i,j) \in E$.}
\end{itemize}

Let us define an algorithm (call it $DirectedMove$), whose parameters are (the number of) the source peg, (the number of) the target peg, and the number of discs to move. The algorithm produces the sequence of moves that moves the discs from the source to the target.
\\\\
\begin{algorithm}[DirectedMove]
  Algorithm DirectedMove(i,j,n) \{\\
\>  \If n=0 exit;\\
\>  k := 6-i-j;\\
\>  \If $(i,j) \in E(G)$ \{\\
\>\>  DirectedMove(i,k,n-1);\\
\>\>     Move disc n from i to j;\\
\>\>      DirectedMove(k,j,n-1);\\
\>      \} \Else \{\\
\>\>      DirectedMove(i,j,n-1);\\
\>\>      Move disc n from i to k;\\
\>\>      DirectedMove(j,i,n-1);\\
\>\>      Move disc n from k to j;\\
\>\>      DirectedMove(i,j,n-1);\\
\>       \}\\
    \}\\
\end{algorithm}    

\subsection{Correctness of the algorithm}{
  \begin{lemma}
    For each $n \geq 0$
    and for each $1 \leq i,j \leq 3$, ($i \neq j$), the algorithm
    DirectedMove moves n disks from i to j correctly.
  \end{lemma}
  \begin{proof}
    By induction on $n$.
    If $n = 0$ the algorithm does nothing.\\
    Assume that the lemma is correct for any number of discs less than $n$
    ($n > 0$), and prove it for $n$ discs.\\
    Assume that $i$ is the source peg, $j$ is the target.We have two
    possibilities:\\
    \begin{enumerate}
    \item $(i,j) \in E$.\\
      The algorithm moves the $n-1$ small discs to the
      auxiliary peg $k$ ($k = 6-i-j$). By induction the algorithm does it correctly, i.e. all discs $\{1, \dots ,n-1\}$ will be on the peg k. Now it moves disc $n$ from peg $i$ to peg $j$, and this is possible because $(i,j) \in E$. And after that, the algorithm moves discs $\{1, \dots ,n-1\}$ from peg $k$ to peg $j$, and by induction it does so correctly.
    \item $(i,j) \notin E$.
      \\
      In this case we cannot make a move from 
      peg i to peg j directly, so, in order to move the disc $n$ from $i$
        to $j$ we must move it first to the peg $k$.\\
        The algorithm makes two moves of the disc $n$: One from $i$ to $k$,
        and another from $k$ to $j$. These moves are possible, since $G$ is
        strongly connected. Also the algorithm makes three moves of the
        discs $\{1, \dots ,n-1\}$ from one peg, to another one, and by
        induction it does so correctly. It is easy to see that each
        time that disc $n$ makes a move from peg $x$ to peg $y$, the discs
        $\{1, \dots ,n-1\}$ are placed on the third peg, so we have the full
        correctness of the algorithm.
      \end{enumerate}
    \end{proof}
    }

\subsection{Optimality of the algorithm}{
  Let $N(i,j,n)$ be the number of moves made by
  algorithm $DirectedMove$ in order to move $n$ discs from peg $i$
  to peg $j$.
  Let $k$ be the third (auxiliary) peg.
    Thus, by the definition of the algorithm, we have:\\
    \begin{equation}
      N(i,j,n) = \left\{
      \begin{array}{ll}
        
      N(i,k,n-1) + N(k,j,n-1) + 1 & \mbox{if $(i,j) \in E$} \\
      2N(i,j,n-1) + N(j,i,n-1) + 2 & \mbox{if $(i,j) \notin E$}
      \end{array} \right.
  \end{equation}
  
  \begin{lemma}
    Let $M(i,j,n)$ be the minimal number of moves required to transfer n discs
    from peg $i$ to peg $j$.
    Then, $M(i,j,0)=0$, and for $n>0$
    \[
    M(i,j,n) = \left\{
      \begin{array}{ll}
        M(i,k,n-1) + M(k,j,n-1) + 1 & \mbox{if $(i,j) \in E$} \\
        2M(i,j,n-1) + M(j,i,n-1) + 2 & \mbox{if $(i,j) \notin E$}
      \end{array} \right.
    \]
  \end{lemma}
  \begin{proof}
    Use induction on $n$.
    If $n=0,1,2$, it is easy to check that the statement is right.
%    If $n=0$, then $M(i,j,0)=0$.\\
%    If $n=1$, then it is clear that
%    \[
%    M(i,j,1) = \left\{
%      \begin{array}{ll}
%        1, & \mbox{if $(i,j) \in E$} \\
%        2, & \mbox{if $(i,j) \notin E$}
%      \end{array} \right.
%    \]
    Consider $n$ discs ($n > 2$) on peg $i$, and we have to transfer them to peg $j$ ($i \neq j$).
    We have two possibilities:
    \begin{enumerate}
      \item $(i,j) \notin E$.\\
        Disc $n$ must move sometime, and its move
        must be to peg $k$ ($k=6-i-j$), because $(i,j) \notin E$.
        So, the discs $\{1, \dots ,n-1\}$ must move first to peg $j$,
        and it takes at least $M(i,j,n-1)$ steps.
        The last position of disc $n$ must be on peg $j$, and the only
        move that can be done here is from peg $k$ to peg $j$
        (because $(i,j) \notin E$).
        So, before the last step of the disc $n$, the discs $\{1, \dots ,n-1\}$
        must be placed on peg $i$, and in order to move them from peg
        $j$ (where they were before) to peg $i$ we must make
        at least $M(j,i,n-1)$ steps.
        After the last move of the disc $n$ we must move all the discs
        $\{1, \dots ,n-1\}$ from peg $i$ to peg $j$, and
        it takes at least $M(i,j,n-1)$ steps.\\
        Now we can conclude, that the number of steps required to move $n$
        discs from peg $i$ to peg $j$ is at least $2M(i,j,n-1) + M(j,i,n-1) + 2$.
        \item $(i,j) \in E$.\\
          Let us look on the packet of discs $\{1, \dots ,n-1\}$. When disc n
          makes a move, all these discs must be on one peg. If sometime the
          packet was on peg $k$ ($k=6-i-j$), we have at least
          $M(i,k,n-1)+M(k,j,n-1)+1$ moves for the transfer.
          If no, it implies the following:
          \begin{enumerate}
            \item Disc $n$ did not make move $i \rightarrow j$, so, edges
              $(i,k)$ and $(k,j)$ are in $E$, and disc $n$ made at least two moves.
            \item The packet of discs $\{1, \dots ,n-1\}$ made at least
              three moves, and the number of moves in the transfer is at
              least $2M(i,j,n-1) + M(j,i,n-1) + 2$.
            \end{enumerate}
            In this case we have to show that
            \[ M(i,k,n-1)+M(k,j,n-1)+1 \leq 2M(i,j,n-1) + M(j,i,n-1) + 2.\]
            $(i,k) \in E$ and $(k,j) \in E$, so, by induction and using the statement that $M(i,j,n-1) \leq M(i,j,n)$ for each $i$,$j$ and for each $n>0$, we have:
            \begin{enumerate}
            \item If $(j,i) \in E$, then
              \[M(i,k,n-1)+M(k,j,n-1)+1 = \]
              \[M(i,j,n-2)+M(j,k,n-2)+1+
              M(k,i,n-2)+M(i,j,n-2)+1+1 =\]
              \[2M(i,j,n-2)+2+M(j,k,n-2)+M(k,i,n-2)+1 \leq \]
              \[  2M(i,j,n-1) + M(j,k,n-2)+M(k,i,n-2)+3 = \]
              \[  2M(i,j,n-1) + M(j,i,n-1)+2.\]
            \item If $(j,i) \notin E$, then $(j,k),(k,i) \in E$ (because $G$
              is strongly connected), and then 
              \[
              M(i,k,n-1)+M(k,j,n-1)+1 = \]
              \[M(i,j,n-2)+M(j,k,n-2)+1+
              M(k,i,n-2)+M(i,j,n-2)+1+1 =\]
              \[2M(i,j,n-2)+\left[ M(j,i,n-3)+M(i,k,n-3)+1 \right]+ \]
              \[ \left[ M(k,j,n-3)+M(j,i,n-3)+1 \right] +3 =\]
              \[2M(i,j,n-2)+\left[ 2M(j,i,n-3)+M(i,j,n-2)+2 \right]+2 \leq \]
              \[2M(i,j,n-2)+ \left[ 2M(j,i,n-2)+M(i,j,n-2)+2 \right]+2 = \]
              \[2M(i,j,n-2)+M(j,i,n-1)+2 \leq \]
              \[2M(i,j,n-1)+M(j,i,n-1)+2.\]
              \end{enumerate}
    \end{enumerate}
    This proves the lemma.
  \end{proof}
  \\
  So, the following claim is proved:
  \begin{claim}
    For each $n \geq 0$ and for each $1 \leq i,j \leq 3$, ($i \neq j$),
    the algorithm $DirectedMove$ moves n discs from peg $i$ to peg $j$
    with minimal possible number of moves. 
  \end{claim}

  }

  \subsection{Calculating the values $N(i,j,n)$.}{
    We turn now to calculate the values of $N(i,j,n)$ as function of $n$ for each
    $1 \leq i,j \leq 3$.\\

    \par Enumerate all pairs $(i,j)$ ($i \neq j$) as follows:\\
    $e_{1}=(1,2)$, $e_{2}=(2,1)$, $e_{3}=(1,3)$, $e_{4}=(3,1)$, $e_{5}=(2,3)$,
    $e_{6}=(3,2)$.\\
    Put $a_{n}^{m}=N(i,j,n)$, where $e_{m}=(i,j)$.\\
    To find the values of each $a_{n}^{m}$, $1 \leq m \leq 6$, we have to
    solve a system of $6$ recursive formulae, each having one of the two forms:
    \\
    $a_{n+1}^{i}=a_{n}^{j}+a_{n}^{k}+1$,
    or $a_{n+1}^{i}=2a_{n}^{i}+a_{n}^{j}+2$,\\
    depending on the existence of the edge $e_{i}$ in $E$.

    It is clear that the system is uniquely defined by the graph $G$.
    Call the system $A$.\\
    \par Consider the generating functions $f_{i}$, $i \leq 6$;
    \[
      f_{i}(x)=\sum_{n=0}^{\infty}a_{n}^{i}x^{n}.
    \]
    \\
    Multiplying the n-th equation in $A$ by $x^{n+1}$ and taking the sum over $n$, we obtain equations of the form
    \[
      \sum_{n=0}^{\infty}a_{n+1}^{i}x^{n+1}=
      x\sum_{n=0}^{\infty}a_{n}^{j}x^{n} +
      x\sum_{n=0}^{\infty}a_{n}^{k}x^{n} +
      x\sum_{n=0}^{\infty}x^{n},
    \]
    or of the form
    \[
      \sum_{n=0}^{\infty}a_{n+1}^{i}x^{n+1}=
      2x\sum_{n=0}^{\infty}a_{n}^{i}x^{n} +
      x\sum_{n=0}^{\infty}a_{n}^{j}x^{n} +
      2x\sum_{n=0}^{\infty}x^{n}.
    \]
Rewriting, we obtain
    \[
      f_{i}(x)-a_{0}^{i} - x f_{j}(x) - x f_{k}(x) = \frac{x}{1-x}
    \]
    or 
    \[
      f_{i}(x)(1-2x)-a_{0}^{i} - x f_{j}(x) = \frac{2x}{1-x}.
    \]
    Here we can substitute value $a_{0}^{i}=0$ for each $1 \leq i \leq 6$.\\
    \par We got a system of $6$ linear equations in the unknowns $f_{j}(x), 1\leq j \leq 6$. The coefficients of the unknowns are from the set $\left\{0,1,-x,1-2x\right\}$, and the free term is $\frac{x}{1-x}$ or $\frac{2x}{1-x}$. \\
    We can find the solution of this system, and get the values of the
    $a_{n}^{i}$, $1 \leq i \leq 6$.\\
    \par For example, for the graph $G=(V,E)$, such that
    $E=\{(1,2),(1,3),(3,1),(2,3)\}=\{e_{1},e_{3},e_{4},e_{5}\}$ ,
    the system of equations is:
    \begin{equation}
      \left(
        \begin{array}{cccccc}
          1 & 0 & {-x} & 0 & 0 & -x\\
          -x & 1-2x & 0 & 0 & 0 & 0\\
          -x & 0 & 1 & 0 & -x & 0\\
          0 & -x & 0 & 1 & 0 & -x\\
          0 & -x & -x & 0 & 1 & 0\\
          0 & 0 & 0 & 0 & -x & 1-2x
        \end{array} \right)
      \left(
        \begin{array}{cccccc}
          f_{1}\\f_{2}\\f_{3}\\f_{4}\\f_{5}\\f_{6}
        \end{array} \right)
      =
      \left(
        \begin{array}{cccccc}
          \frac{x}{1-x}\\
          \frac{2x}{1-x}\\
          \frac{x}{1-x}\\
          \frac{x}{1-x}\\
          \frac{x}{1-x}\\
          \frac{2x}{1-x}
        \end{array} \right)
    \end{equation}
    }
%\pagebreak
\\
In figures 1-4 we show graphicaly all possible (non-isomorphic) graphs $H(V,E)$. Directed edge $<A,B>$ means, that the move from peg A to peg B is permited. An undirected edge means that both directed edges $<A,B>$ and $<B,A>$ are in $E$.
\\
\hrule
\begin{figure}[h]
  \vbox{\center\epsfig {figure=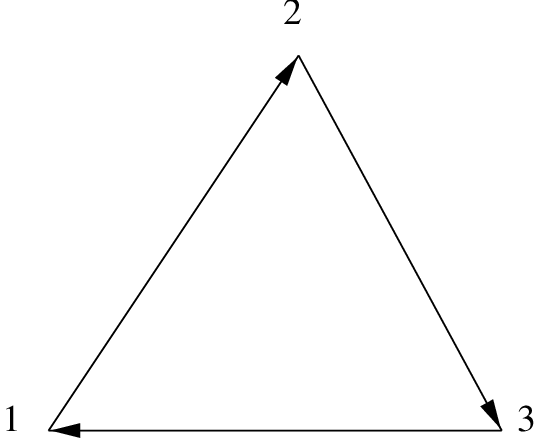,width=2cm,height=2cm,angle=270}}
  \center{Figure 1.}

\end{figure}
Here we deal with the case where only 3 types of moves are permitted.
\[f_{2}=f_{3}=f_{6},\]
\[f_{6}=\frac{x^{2}+2x}{(1-x)(1-2x-2x^{2})}=-\frac{1}{2}\left(\frac{A}{1-x}+\frac{B}{x+\frac{1+\sqrt{3}}{2}}+\frac{C}{x+\frac{1-\sqrt{3}}{2}}\right).\]
\par The result is
\[N(3,2,n)=N(2,1,n)=N(1,3,n)=\frac{2+\sqrt{3}}{2\sqrt{3}}(1+\sqrt{3})^{n}-\frac{2-\sqrt{3}}{2\sqrt{3}}(1-\sqrt{3})^{n}-1.\]

\[f_{5}=f_{1}=f_{4},\]
\[f_{5}=\frac{x(1+2x)}{(1-x)(1-2x-2x^{2})}=-\frac{1}{2}\left(\frac{A}{1-x}+\frac{B}{x+\frac{1+\sqrt{3}}{2}}+\frac{C}{x+\frac{1-\sqrt{3}}{2}}\right).\]
\par The result is
\[N(2,3,n)=N(1,2,n)=N(3,1,n)=\frac{1+\sqrt{3}}{2\sqrt{3}}(1+\sqrt{3})^{n}-\frac{1-\sqrt{3}}{2\sqrt{3}}(1-\sqrt{3})^{n}-1.\]
\hrule
%\pagebreak
%%%%%%%%%%%%%%%%%%%%%%%%%%%%%%%%%%
\hrule
\begin{figure}[h]
  \vbox{\center\epsfig {figure=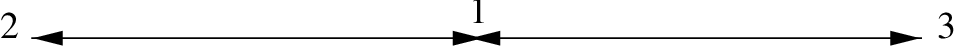,width=.3cm,height=5cm,angle=270}}
  \center{Figure 2.}

\end{figure}
\[f_{5}=f_{6},\]
\[f_{5}=\frac{2x}{(1-x)(1-3x)}=\frac{A}{1-x}+\frac{B}{1-3x}.\]
\par The result is
\[N(2,3,n)=N(3,2,n)=3^{n}-1.\]
\[f_{1}=f_{2}=f_{3}=f_{4},\]
\[f_{4}=\frac{x}{(1-x)(1-3x)}=\frac{A}{1-x}+\frac{B}{1-3x}.\]
\par The result is
\[N(1,2,n)=N(2,1,n)=N(1,3,n)=N(3,1,n)=\frac{3^{n}-1}{2}.\]
\hrule
%%%%%%%%%%%%%%%%%%%%%%%%%%%%%%%%%%%
\hrule
\begin{figure}[h]
  \vbox{\center\epsfig {figure=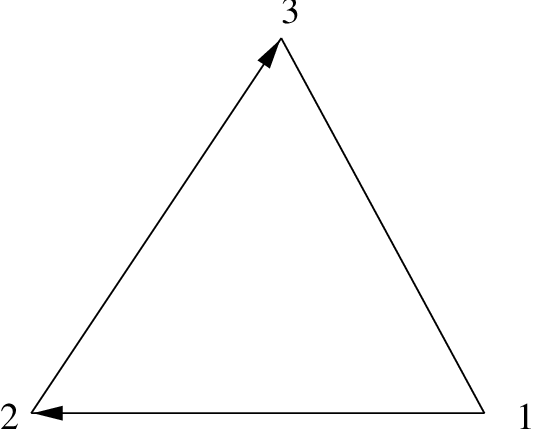,width=2cm,height=2cm,angle=270}}
  \center{Figure 3.}
\end{figure}
\[f_{1}=f_{5},\]
\[f_{5}=\frac{x(1+2x)}{(1-x)(-4x^{2}-x+1)}=\frac{A}{1-x}+\frac{B}{x+\frac{1+\sqrt{17}}{8}}+\frac{C}{x+\frac{1-\sqrt{17}}{8}}.\]
\par The result is
\[N(2,3,n)=N(1,2,n)=-3/4-\frac{11-3\sqrt{17}}{8\sqrt{17}}\left(\frac{1-\sqrt{17}}{2}\right)^{n}+\frac{11+3\sqrt{17}}{8\sqrt{17}}\left(\frac{1+\sqrt{17}}{2}\right)^{n}.\]

\[f_{4}=\frac{x(2x^{2}+3x+1)}{(1-x)(1-x-4x^{2})}=\frac{A}{1-x}+\frac{B}{x+\frac{1+\sqrt{17}}{8}}+\frac{C}{x+\frac{1-\sqrt{17}}{8}}\]
\par The result is

\[N(3,1,n) = \left\{
        \begin{array}{ll}
          
          \frac{1}{2}\left(1-3+\frac{4+\sqrt{17}}{\sqrt{17}}-\frac{4-\sqrt{17}}{\sqrt{17}}\right)=0 & n=0\\
          \frac{1}{2}\left(-3+\frac{4+\sqrt{17}}{\sqrt{17}}\left(\frac{1+\sqrt{17}}{2}\right)^{n}-\frac{4-\sqrt{17}}{\sqrt{17}}\left(\frac{1-\sqrt{17}}{2}\right)^{n}\right) & n\geq1
        \end{array}
        \right.\]
%    \end{equation}

\[f_{2}=f_{6},\]    
\[f_{6}=\frac{x(3x+2)}{(1-x)(1-x-4x^{2})}.\]
\par The result is
\[N(3,2,n)=N(2,1,n)=-5/4-\frac{21-5\sqrt{17}}{8\sqrt{17}}\left(\frac{1-\sqrt{17}}{2}\right)^{n}+\frac{21+5\sqrt{17}}{8\sqrt{17}}\left(\frac{1+\sqrt{17}}{2}\right)^{n}.\]

\[f_{3}=\frac{x(x+1)}{(1-x)(-4x^{2}-x+1)}=\frac{A}{1-x}+\frac{B}{x+\frac{1+\sqrt{17}}{8}}+\frac{C}{x+\frac{1-\sqrt{17}}{8}}.\]
\par The result is
\[N(1,3,n)=-1/2-\frac{5-\sqrt{17}}{4\sqrt{17}}\left(\frac{1-\sqrt{17}}{2}\right)^{n}+\frac{5+\sqrt{17}}{4\sqrt{17}}\left(\frac{1+\sqrt{17}}{2}\right)^{n}.\]
\hrule

%%%%%%%%%%%%%%%%%%%%%%%%%%%%%
%\pagebreak
\hrule
\begin{figure}[h]
  \vbox{\center\epsfig {figure=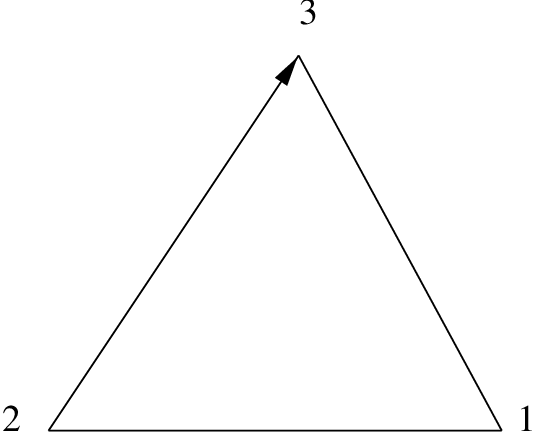,width=2cm,height=2cm,angle=270}}
  \center{Figure 4.}

\end{figure}
\par Here we have more complex generating functions, which denominator's multiplier is $2x^{3}-4x^{2}-x+1$. The exact solution of the equation is more complex, but it is easy to check, that the greatest real root is $\sim 2.12$. In all solutions of kind $N(i,j,n)$ the major element will be $\sim (2.12)^{n}$, and this will be an order of the solution.
\\
\hrule
%%%%%%%%%%%%%%%%%%%%%%%%%%%
%\pagebreak
  \section{Solution of problem 1.3 for $C=1$}
    Consider $n \in \N$. Let $\zeta_{n}$ be the sequence of moves that transfers $n$ discs from peg $i$ to peg $j$ with auxiliary peg $k$. The definition is recursive:

    \begin{itemize}
      \item $n=0$: the state does not change.
      \item $n=1$: move the disc from $i$ to $j$.
      \item $n \geq 2$: move the $n-2$ smallest discs from $i$ to $k$ by $\zeta_{n-2}$
        , move by two steps discs $n-1$ and $n$ from $i$ to $j$,
        and then move the $n-2$ small discs from $k$ to $j$ by $\zeta_{n-2}$.
    \end{itemize}

  {\bf Note:} It is clear that there are initial states for which $\zeta_{n}$
  is legal, and there are states for which $\zeta_{n}$ is illegal.
  It is easy to show that $\zeta_{n}$ is legal for the standard initial state.
  Denote by $b_{n}$ the length of $\zeta_{n}$. \\\\
  Clearly, $b_{0}=0$, $b_{1}=1$ and $b_{n}=2b_{n-2}+2$ for $b \geq 2$.
  It is easy to show by induction that $b_{n}>b_{n-1}$ for $n \geq 1$.\\
  In Theorem \ref{t1} we will show that $b_{n}$ is a minimal number of moves
  required to transfer n disks from one peg to another.\\
    \par {Given a legal state X of $n$ discs on three pegs, denote by $X_{-n}$ the
    state (with $n-1$ disks) obtained by removing disc $n$ from the state.}
  \begin{lemma} \label{l1}
    Let $X$,$Y$ be two legal states on three pegs. Assume that by some sequence of
    moves $\alpha$ we have passed from $X$ to $Y$.If the length of the sequence is $|\alpha|$,
    there exists sequence of moves $\beta$ from $X_{-n}$ to $Y_{-n}$ of length $|\beta|$, such that \[|\alpha|=|\beta|+k\] where $k$ is the number of moves that the disc $n$ does in $\alpha$.
  \end{lemma}
  \begin{proof}
    Let $\beta$ be the sequence obtained from $\alpha$ by removing all moves of
    disc $n$. It is easy to show by induction that $\beta$ is legal sequence
    that passes from $X_{-n}$ to $Y_{-n}$.
  \end{proof}
  
  \begin{theorem} \label{t1}
    Assume we have $n$ discs, and $X$ is a legal state with all discs on a single peg. Any legal transfer of all discs to another peg requires at least $b_{n}$ moves.
  \end{theorem}
  \begin{proof}
    Disc $n$ must move sometime, so, if it makes the first move from peg
    $1$ to peg $2$, the discs $\{1, \dots ,n-2\}$ must be in that time on 
    peg $3$, in order to make the move legal.
    So, we have moved all the $n-2$ small discs from one peg to another, that
    by induction and by Lemma \ref{l1} is done by at least $b_{n-2}$ steps.
    And after the last step of the disc $n$ we must have the discs
    $\{1, \dots ,n-2\}$ gathered on one peg (not the $target$), so to move them
    to the target we must make (by induction and Lemma \ref{l1}) at least
    $b_{n-2}$ steps. Also, disks $n$ and $n-1$ must move at least once,
    so we have at least $2b_{n-2}+2=b_{n}$ steps. This proves the theorem.
  \end{proof}
  \begin{lemma} \label{l2}
    If $X$ is a legal state with all $n$ discs on one peg, $Y$ a legal
    state with disc $n$ on one peg, and the other
    $n-1$ discs on another peg, then passing from $X$ to $Y$ requires
    at least $b_{n-1}$ steps. 
  \end{lemma}
  \begin{proof}
    If disc $n$ has not moved from the initial peg, then, by Theorem \ref{t1}
    we have at least $b_{n-1}$ steps.\\
    If it has moved, then let us consider the discs $\{1, \dots ,n-2\}$.
    This packet must move from peg to peg at least once, because we have to
    move the large disc.If the packet makes one move, all the discs
    $\{1, \dots ,n-1\}$ move from one peg to another, and this by Theorem
    \ref{t1} requires at least $b_{n-1}$ steps.
    If the packet makes more than one move, then the number of steps is 
    at least
    \[ 2b_{n-2}+1=b_{n}-1 \geq b_{n-1}.\]
    The lemma is proved.
  \end{proof}
  
  \begin{lemma} \label{df1}
    Let $d_{n}=x_{n}-y_{n}$, where
    \[x_{n}=2b_{n-1}+1, y_{n}=3b_{n-k}+2k, b_{n}=2b_{n-k}+k.\]
    Then $d_{n}=2d_{n-k}+k-1$.
  \end{lemma}
  \begin{proof}
    By the definition of $b_{n}$
    \[x_{n}=2b_{n-1}+1=2(2b_{n-1-k}+k)+1=4b_{n-1-k}+2k+1,\]
    \[y_{n}=3b_{n-k}+2k=3(2b_{n-2k}+k)+2k=6b_{n-2k}+5k.\]
    Therefore:
    \center $x_{n}-y_{n}=2b_{n-1}+1-3b_{n-k}-2k=$
    \center $=4b_{n-1-k}-6b_{n-2k}-3k+1=$
    \center $=2(2b_{n-1-k}-3b_{n-2k}-2k+1)-2(-2k+1)-3k+1=2(x_{n-k}-y_{n-k})+k-1$
    \[d_{n}=x_{n}-y_{n}=2d_{n-k}+k-1.\]
  \end{proof}
  {\bf Conclusion:} By Lemma \ref{df1}, if $b_{0}=0$, $b_{1}=1$ then $2b_{n-1}+1 < 3b_{n-2}+4$ for each $n \geq 2$ (it is easy to prove by induction, that $d_{n}<0$ in this case).
  
  \begin{defn}
    Let us call the pegs $initial$, $target$ and $auxiliary$ arbitrarily. 
    \begin{enumerate}
    \item{
        A state is a $\lambda$-state if disc $n$ is
        on the initial peg, disc $n-1$ is on some
        other peg, and the discs $\{1, \dots ,n-2\}$ are placed on the
        third peg in any legal order.}
    \item{A state is a $\lambda '$-state if the initial peg is empty, disc $n$ is placed over disc $n-1$ on other peg, and discs $\{1, \dots ,n-2\}$ are on the third peg in any legal order.}
    \end{enumerate}
  \end{defn}
  \begin{lemma} \label{ll1}
  Consider we start from a $\lambda '$-state, and end with the standard state on the target peg, the sequence of moves is of length at least $2b_{n-2}+2$.
  \end{lemma}
  \begin{proof}
    Prove it by induction on the number of $\lambda$-states we encounter in the sequence. 
    \begin{enumerate}
    \item Assume the sequence leads to no $\lambda$-states. Disc $n$ must
      make a move, and so, by our assumption, it must move to the peg where now discs $\{1, \dots ,n-2\}$ are placed (otherwise we arrive at a $\lambda$-state). Hence we must move discs $\{1, \dots ,n-2\}$ from that peg to another, and that, by Theorem \ref{t1}, will take at least $b_{n-2}$ steps.
      After moving disc $n$ we must gather all the discs on some peg, which will take at least $b_{n-2}+1$ steps (by Lemma \ref{l2}).\\
      Thus, the whole transfer takes at least $b_{n-2}+1+b_{n-2}+1=2b_{n-2}+2$ steps.
    \item
      Let us make now the induction step.
      Assume the sequence yields k $\lambda$-states. Consider the
      first move of disc $n$. If the move is to the peg where
      discs $\{1, \dots ,n-2\}$ are, we have the same
      situation as in (a).\\
      Otherwise, after the first move we have a $\lambda$-state.
      Now we have two possibilities:
      \begin{enumerate}
      \item If disc $n$ moves to the peg where discs $\{1, \dots ,n-2\}$ are, then as before we show that the sequence has taken at least $b_{n-2}$ steps to move the $n-2$ small discs to another peg, and at least $b_{n-2}+1$ steps to gather all the discs on one peg.
      \item If the disc moves to the peg where it was before in the
        $\lambda '$-state, then either it is placed over disc $n-1$,
        and then we can apply induction, or it is not, and we have to do at least $b_{n-2}$ steps to move discs $\{1, \dots ,n-2\}$ to another peg, and at least $b_{n-1}$ step to gather all together.
      \end{enumerate}
      It follows that, in any case, we have at least $2b_{n-2}+2$ steps for the transfer.
    \end{enumerate}
    The lemma is proved.
  \end{proof}

  \begin{theorem} \label{t2}
    Let the initial state $X$ be standard, and the terminal state $Y$, be standard too, with the discs on another peg.
    Then, any transfer from $X$ to $Y$ takes at least $2b_{n-1}+1$ steps. 
  \end{theorem}
  \begin{proof}
    Consider the first move of disc $n$.
    \begin{enumerate}
    \item{
      If the move is to an empty peg, then we had to move all the discs
      $\{1, \dots ,n-1\}$ to another peg, that have taken, by Theorem
      \ref{t1},at least $b_{n-1}$ steps. In order to gather all discs
      on one peg, by Lemma \ref{l2}, we have to make at least
      $b_{n-1}$ steps.We have made one move of disc $n$. Altogether we have
      at least $2b_{n-1}+1$ steps to the transaction.}
    \item{
      Assume the first move of disc $n$ is to the peg, where disc
      $n-1$ is placed (in other words, we came to the $\lambda '$-state). Then, by Lemma \ref{l2}, before this move we made
      at least $b_{n-1}+1$ steps, and by Lemma \ref{ll1} the transfer to the standard state on the target peg has taken at least $2b_{n-2}+2$ steps.\\
      
      Now we see that the total length of the sequence is at least $b_{n-2}+2+2b_{n-2}+2=3b_{n-2}+4$, and by the conclusion after Lemma \ref{df1} we can see that for each $n \geq 2$ we have $2b_{n-1}+1 < 3b_{n-2}+4$.}
    \end{enumerate}
    This proves the theorem.
  \end{proof}
    
%%%%%%%%%%%%%%%%%%%%%%%%%%%

  By Theorem \ref{t2} we can claim the following:
  \begin{claim}
    The number of steps in the optimal solution of the problem with $C=1$ can
    be calculated by the following system:
    \begin{equation}
      \left\{
        \begin{array}{ll}
          a_{n}=2b_{n-1}+1,\\
          b_{n}=2b_{n-2}+2,
        \end{array} \right.
    \end{equation}
    where $a_{0}=0$, $b_{0}=0$, $b_{1}=1$.\\
    Explicitly:
    \begin{equation}
      a_{n}=\left(\frac{3+2\sqrt{2}}{2}\right)(\sqrt{2})^{n}+
      \left(\frac{3-2\sqrt{2}}{2}\right)(-\sqrt{2})^{n}-3
    \end{equation}
  \end{claim}

%\pagebreak
\section{Unsolved problems}
This part of the paper is dedicated to an elaboration on problem \ref{pmain}, where $C\geq2$.

{\bf Conjecture.} The solution of the main problem is
\begin{equation}
\left\{
        \begin{array}{ll}
          a_{n}=2b_{n-1}+1\\
          b_{n}=2b_{n-C-1}+C+1
        \end{array} \right.
\end{equation}
where C is a distance of the current model.
\\
We were unable to prove the conjecture. The following is intended to shed some light on it.
\\\\
\underline{\bf{\large The lower bound for the shortest sequence length}}
\vspace{.5cm}
\par{
  The first stage of the proof, that we thought was to try to claim that there is no shortest sequence of moves than $a_{n}$. Let us try to check the sequence of moves $q_{n}$ as follows :
\\  
  \begin{enumerate}
    \item{Move $n-C-1$ discs to the target peg.}
    \item{Move c+1 discs directly from the source peg to the auxiliary peg.}
    \item{Move $n-C-1$ discs from the target peg to the source peg.}
    \item{Move c+1 discs directly from the auxiliary peg to the target peg.}
    \item{Move $n-C-1$ discs from the source peg to the target peg.}  
  \end{enumerate}
The steps, that are interesting for us are 1, 3 and 5. For example, we can decide that in 1 and 3 we want to move discs as quick as possible and the fifth step must recursively call the same procedure, but with $n-C-1$ discs. Let us check this case.
\begin{claim}
  Given is next system of equations
  \[
  \left\{
    \begin{array}{lll}
          x_{n}=2b_{n-k}+x_{n-k}+2k\\
          y_{n}=2b_{n-1}+1\\
          b_{n}=2b_{n-k}+k
    \end{array}
  \right.
  \]
  where $k=C+1$, then for $n \geq k$ :
  \[x_{n}-x_{n-k} \geq y_{n}-y_{n-k}.\]
\end{claim}
\begin{proof}
  \[x_{n}-x_{n-k}=2b_{n-k}+2k,\]
  \[y_{n-k}=2b_{n-k-1}+1, y_{n}-y_{n-k}=2(b_{n-1}-b_{n-k-1}).\]
  It is easy to show, that $b_{n}$ is non-decreasing sequence, so
  \[b_{n-k}+k \geq b_{n-k-1}+k = b_{n-1}-b_{n-k-1}.\]
  now we have
  \[2(b_{n-k}+k) \geq 2(b_{n-1}-b_{n-k-1}).\]
  and this proves that
  \[x_{n}-x_{n-k} \geq y_{n}-y_{n-k}.\]
\end{proof}
In the previous claim we think, that $x_{n}$ and $y_{n}$ supposed to be minimal sequences that we need, so if $x_{i} = y_{i}, 1 \leq i \leq k$ then we get that the sequence $y_{n}$ is not minimal.
\\
Another way is to say, that all 3 steps must be as short as possible. We can prove, that the shortest sequence of moves from one legal state to another is at least as long as $\zeta_{n}$. But the combination of all 3 shortests steps cannot give us the the right algorithm for moving from a standard state to a standard state on another peg. Unfortunately, for $C\geq2$ we got that the length of  the sequence generated by $q_{n}$ is less than the length of $a_{n}$, so we cannot claim that there is no algorithm that returns to us the sequence of moves that lies between the lengths of $q_{n}$ and $a_{n}$.}

\vspace{.5cm}
\hspace{-.5cm}\underline{\bf{\large Symmetry of one of the shortest sequences of moves}}
\vspace{.5cm}
\par{
  \begin{defn}
    Let $\ga$ be sequence of moves with $|\ga|=n$. $\ga$ is a symmetric sequence of moves if the following conditions are satisfied:
    \begin{itemize}
      \item{the i-th move was done with disc $j$, then (n+1-i)-th move would be done also with disc $j$}
      \item{the i-th move was $start \rightarrow target$ iff (n+1-i)-th move would be done also $start \rightarrow target$}
      \item{the i-th move was $target \rightarrow start$ iff (n+1-i)-th move would be done also $target \rightarrow start$} 
      \item{the i-th move was $aux \rightarrow target$ iff (n+1-i)-th move would be $start \rightarrow aux$}
      \item{the i-th move was $aux \rightarrow start$ iff (n+1-i)-th move would be $target \rightarrow aux$}
    \end{itemize}
  \end{defn}

For example the following sequence of moves :
\[s \rightarrow t, s \rightarrow a, s \rightarrow a, t \rightarrow a, s \rightarrow t, a \rightarrow s, a \rightarrow t, a \rightarrow t, s \rightarrow t\]
is the symmetric sequence that transfers 4 discs from peg $s$ to peg $t$ using intermediate peg $a$ ($C=1$). We can see that this sequence is a shortest possible.
\\\\
We noted, that in any model there is at least one shortest symmetric sequence of moves. If this claim is proved, then we can formulate the next claims :
\begin{claim}
  If $\alpha_{n}$ is a symmetric shortest sequence of moves in model M, then $|\alpha_{n}|$ is odd.
\end{claim}
\begin{claim}
  If $\alpha_{n}$ is a symmetric shortest sequence of moves in model M, then $|\alpha_{n}|=a_{n}$.
\end{claim}
We have proved these claims. The main points of the proof of the last claim are:
    \begin{enumerate}
      \item{Use the previous claim}
      \item{If the length is odd, then the middle move must be made by the largest disc from the source peg to the target peg.}
      \item{If step 2 is true, then the shortest sequence is of length $a_{n}$.}
    \end{enumerate}

    Unfortunately, the symmetry of the arbitrary model is not proved.
}

\pagebreak

\vspace{.5cm}
{\large \bf References.}
\begin{enumerate}
\item[[1]] D.Wood, The Towers of Brahma and Hanoi revisited ,
           submitted to ``{\em Journal of Recreational mathematics}'', Vol.14(1), 1981-1982
\end{enumerate}

\end{document}